\documentclass[journal=apchd5,manuscript=article,layout=traditional]{achemso}

\usepackage{graphicx}
\usepackage[intlimits]{amsmath}
\usepackage{amssymb}
\usepackage{amsthm}
\usepackage{booktabs}
\usepackage{multirow}
\usepackage{natbib}
\usepackage{geometry}
\usepackage{color}
\usepackage{bm}

\definecolor{olivegreen}{rgb}{0,0.5,0}

\author{R. Carmina Monreal}
\affiliation{Departamento de F\'{\i}sica Te\'orica de la Materia Condensada C5 and Condensed Matter Physics Center (IFIMAC), Universidad Aut\'onoma de Madrid, E-28049 Madrid, Spain}
\email{r.c.monreal@uam.es}
\author{S. Peter Apell}
\affiliation{Department of Physics and Gothenburg Physics Centre, Chalmers University of Technology, SE-412 96 G\"oteborg, Sweden}
\author{Tomasz J. Antosiewicz}
\email{tomasz.antosiewicz@uw.edu.pl}
\affiliation{Centre of New Technologies, University of Warsaw, Banacha 2c, 02-097 Warsaw, Poland}
\alsoaffiliation{Department of Physics and Gothenburg Physics Centre, Chalmers University of Technology, SE-412 96 G\"oteborg, Sweden}

\title{Quantum-Size Effects in the Visible Photoluminescence of Colloidal ZnO Quantum Dots: A Theoretical Analysis}

\keywords{Quantum dots, photoluminescence, zinc oxide}

\begin{document}

\newpage

\begin{abstract}
ZnO has been known long since to be a highly efficient luminescent material. In the last years, the experimental investigation of the
luminescent properties of colloidal ZnO nanocrystals in the nanometer range of sizes has attracted a lot of of interest for their potential applications in
light-emitting diodes and other optical devices and in this work we approach the problem from a theoretical perspective.
 Here, we  develop a simple theory for the green photoluminescence of ZnO quantum dots (QDs) 
 that allows us to understand and rationalize several experimental findings on fundamental grounds. 
We calculate the spectrum of light emitted in the radiative recombination of a conduction band electron with a deeply trapped hole 
and find that the experimental behavior of this emission band with particle size can be understood in terms of quantum size effects of the electronic states and their overlap 
with the deep hole.
We focus the comparison of our results on detailed experiments performed for colloidal ZnO nanoparticles
in ethanol and find that the experimental evolution of the luminescent signal with particle size
 at room temperature can be better reproduced by assuming the deep hole to be localized at the surface of the nanoparticles. 
However, the experimental behavior of the intensity and decay time of the signal with temperature can be rationalized in terms of 
holes predominantly trapped near the center of the nanoparticles at low temperatures being transferred to surface defects at room temperature.
Furthermore, the calculated values of the radiative lifetimes are comparable to the experimental values of the decay time of the 
visible emission signal.
We also study the visible emission band as a function of the number of electrons in the conduction band of the nanoparticle, 
finding a pronounced dependence of the radiative lifetime but a weak dependence of energetic position of the maximum intensity.
\end{abstract}

\maketitle

\newpage

\section*{Introduction}

ZnO has been known long since to be a highly efficient fluorescent material \cite{JOSA_11_939_shrader, SolStatePhys_8_191_heiland} and as such 
proposed as an excellent candidate for light-emitting and optoelectronic devices \cite{MaterSciEngB_80_383_look}. Many studies done in the past using
single crystals and powders \cite{JPC_88_5556_anpo, JApplPhys_79_7083_vanheusden} showed that when ZnO is excited with UV light across the bandgap, two emission bands appear. The narrow band, known as the exciton band, is at an energy close to the incident energy and originates from the radiative recombination of a
photoexcited electron with a valence band hole. The broad band is in the visible and is called the green luminescence band because of its color. 
It is associated with the existence of intrinsic lattice defects, vacancies or impurities \cite{SolStatePhys_8_191_heiland, JPC_88_5556_anpo, PRL_23_579_dingle, JChemPhys_22_250_kroger} 
that can trap charge carriers and for this reason it is also called the defect luminescence band.

In the last years the luminescent properties of colloidal ZnO quantum dots (QDs) in the nanometer range of sizes have been investigated. The two bands 
also appear in this case and their relative intensity is extremely dependent on whether the nanoparticles are in an atmosphere with or without the presence of oxygen \cite{ChemPhysLett_122_507_koch, JPhysChem_91_3789_bahnemann, JPCB_104_1715_vanDijken, JLumin_90_123_vanDijken, JPCB_104_4355_vanDijken, JPCC_114_220_stroyuk, JPCC_115_21635_yamamoto,
JPCC_116_20633_cohn, JACS_123_11651_shim, JPCL_4_3024_faucheux}. Under aerobic conditions the visible band is prominent while the exciton band is a weak feature. However, under anaerobic conditions and upon irradiation with UV light, the visible emission band quenches as the exciton band increases in intensity 
\cite{ChemPhysLett_122_507_koch, JPhysChem_91_3789_bahnemann, JPCB_104_1715_vanDijken, JLumin_90_123_vanDijken, JPCB_104_4355_vanDijken, JPCC_114_220_stroyuk, JPCC_115_21635_yamamoto,
JPCC_116_20633_cohn}. Furthermore, an infrared absorption band develops due to the accumulation of electrons in the conduction band that leads to the formation of
 a local surface plasmon resonance in the nanoparticle \cite{JACS_123_11651_shim, JPCL_4_3024_faucheux, ACSNano_8_1065_schimpf}. 
Admission of oxygen provokes the opposite effect: the luminescent exciton emission disappears as the visible emission band recovers in intensity
\cite{JPCB_104_4355_vanDijken, JPCC_114_220_stroyuk}
and the infrared absorption is also quenched \cite{JPCL_4_3024_faucheux}. 

There are few experiments analyzing in detail the size dependence of the luminescent properties of ZnO QDs 
\cite{ChemPhysLett_122_507_koch,JPhysChem_91_3789_bahnemann, JPCB_104_1715_vanDijken, JLumin_90_123_vanDijken}. Such dependence is interesting since for these nanocrystals the size-quantization of
the energy levels of electrons and holes is expected to determine their emission properties. 
 The most extensive experiments were conducted in air for suspensions of ZnO nanoparticles in 2-propanol or ethanol in 
 \cite {JPCB_104_1715_vanDijken, JLumin_90_123_vanDijken}, where the intensity and the position of the visible band, as well as the decay time of the luminescent signal were monitored as a function of size at room temperature. It was found that, as the nanoparticle size increases, the intensity of the band decreases,  its position red shifts 
and the decay time increases, in accordance with previous observations \cite{ChemPhysLett_122_507_koch,JPhysChem_91_3789_bahnemann}. 
Also, the intensity and the decay time are very dependent on temperature: 
the intensity decreases by a factor
of 5 and the decay time increases by approximately the same factor as the temperature increases from 4 K to room temperature \cite{JPCB_104_1715_vanDijken}.
Up to our knowledge, the evolution of the visible luminescence signal with nanoparticle size has not yet been studied under anaerobic conditions with the same detail, probably
because of its low intensity. Even for a given size, there are contradictory experiments concerning the shift of this band as the crystal 
becomes charged: while a blue shift is reported in \cite{JPCB_104_4355_vanDijken, JPCC_115_21635_yamamoto, JPCL_4_3024_faucheux}, other Authors observed essentially
no shift \cite{JPCC_114_220_stroyuk}.

The current theoretical investigation of colloidal QDs is focused on calculations of the electronic band structure of 
clusters with sizes smaller than 3 nm in diameter,
with and without ligands at the surface, and at different levels of 
sophistication (we refer the reader to references \cite{AccChemRes_49_2127_kilina, JPCL_8_5209_giansante} and references therein)
but such calculations cannot be extended to larger systems nowadays. In this work we attack the problem from a different perspective and 
 develop a simple theory for the defect photoluminescence of  QDs 
 that allows us to understand and rationalize several experimental findings on fundamental grounds.
 Using Fermi's golden rule together with a simple model for the quantized electronic states, we analyze theoretically the characteristics of the visible emission 
 band of ZnO QDs and their evolution with particle size.
We focus on comparing our results with the detailed experiments of van Dijken et al. \cite{JPCB_104_1715_vanDijken, JLumin_90_123_vanDijken} 
performed for ZnO nanoparticles in ethanol under aerobic conditions.
The behavior of the spectral characteristics with particle size can be understood in terms of quantum size effects and the 
experiments at room temperature are reproduced more accurately by assuming the hole to be localized at the surface of the nanoparticle. 
 Furthermore, the theoretical values of the radiative lifetime are comparable to the experimental values of the decay time.
Based on our results, we suggest that the anomalous dependence of the decay time with temperature found by van Dijken et al. \cite{JPCB_104_1715_vanDijken} can be the effect of an increasing probability for the holes to be trapped at the surface with increasing temperature. 
We also study the visible emission band as a function of the number of electrons in the nanoparticle finding a pronounced dependence of the radiative lifetime.
We conclude that a better experimental characterization 
of the distribution of traps and the charge of the nanoparticles is essential to fully understand and control their emission properties.

\section*{Theory}
Although there has been some controversy concerning the origin of the mechanism of the visible luminescence of ZnO, 
 whether it is due to the radiative recombination of (i) a delocalized  electron with a deeply trapped hole or 
(ii) a deeply trapped electron with a delocalized  hole
 \cite{JPCB_104_1715_vanDijken, JPCC_114_220_stroyuk, PCCP_2016_camarda}, it is now well established that mechanism (i) is dominant
\cite{JLumin_90_123_vanDijken, JPCC_116_20633_cohn, JPCL_4_3024_faucheux} and in this work we adhere to this picture. 

Our system consists of ZnO spheres of radius $R$ is a medium characterized by its dielectric permittivity $\epsilon_m$.
When the sphere is illuminated with light of energy $\hbar \omega_{in}$ across the bandgap,
a photon is absorbed  
exciting an electron to the conduction band and leaving a hole in the valence band. Some of the holes can be transferred very quickly to hole traps 
while some of the electrons can be scavenged by 
oxygen adsorbed at the nanoparticle surface, especially for particles in air, before light emission takes place.  Therefore the number of electrons $N_e$ and the number of holes
$N_h$ participating in the subsequent luminescent process are in general different. Since the charge and the number and distribution of hole traps are
not well characterized experimentally, we need some simple but reasonable assumptions for $N_e$ and $N_h$. 
Given that the exciton emission band is very weak when the nanoparticles are in air, we assume that 
all the created holes are transferred to hole traps. Then $N_h$ equals the number of absorbed photons 
(per photon incident onto the sphere):  
 $N_h=\frac{\sigma_{abs}(\omega_{in}, R)}{2 \pi R^2}$, where 
$\sigma_{abs}(\omega_{in},R)$ is the absorption cross section of a ZnO sphere of radius $R$. 
With respect to the number of electrons in the conduction band we assume either of two cases in which (a) $N_e$ is independent of size or (b) the electronic density
$n_e=3 N_e/(4 \pi R^3)$ is independent of size. In all cases we consider the electronic system to be in its ground state in the conduction band 
when light emission occurs. We work at 
 $T=0$ K but our results are also valid at room temperature since the excitation energies of the QDs are 
 typically of $\simeq 0.1$ eV and therefore larger than the thermal energy at room temperature. 

To study steady-state luminescence, we calculate the probability per unit time 
that a conduction band electron fills a deep hole with emission of a photon of frequency $\omega$ using Fermi's golden rule 
 \begin{equation}
\frac{1}{\tau_{ph}}=\frac{2 \pi}{\hbar}\sum_{photons} \sum_{i} f(e_i) |M_{h,i}|^2 \delta(e_h-e_i+\hbar \omega),
\label{golden0}
\end{equation} 
where $\sum_{photons}$ indicates summation over all possible photon states, $\sum_i$ indicates summation over all possible states $i$  of a conduction electron, $f(e_i)$ is the Fermi factor giving the occupancy of state $i$ of energy $e_i$, the final state $h$ is the trapped-hole state of energy $e_h$ 
and the $\delta$-function expresses energy conservation. 
The matrix elements $M_{h,i}$ for the transition are
\begin{equation}
M_{h,i}=\frac{e}{2 m_{e}^{*} c} \langle \Psi_h|\hat {p}\cdot \vec{A}_{\mu}^{*}+\vec{A}_{\mu}^{*}\cdot \hat {p}|\Psi_i \rangle .
\label{M0}
\end{equation} 

In eq.~(\ref{M0}) $e$ and $m_{e}^{*}$ are the charge and effective mass, respectively, of a conduction band electron, $c$ is the speed of light,
 $|\Psi_i\rangle $ and $|\Psi_h\rangle $ 
are the electronic initial and the hole final states, respectively, $\hat p=-i \hbar \vec{\nabla}$ is the momentum operator 
and $\vec{A}_{\mu}$ is the vector potential of the electromagnetic field with polarization $\mu$.
In the Coulomb gauge $\vec{A}_{\mu}^{*}=i \frac{c}{\omega}\vec{E}_{\mu}^{*}$, $\vec{E}_{\mu}$ being the electric field vector, 
so that $M_{i,h}$ is rewritten as
\begin{equation}
M_{h,i}=\frac{e \hbar}{2 m_{e}^{*} \omega} \langle \Psi_h|\vec{\nabla}\cdot \vec{E}_{\mu}^{*}+2\vec{E}_{\mu}^{*}\cdot \vec{\nabla}|\Psi_i \rangle.
\label{M'}
\end{equation} 
The electric field $\vec{E}_{\mu}$ is defined except for a normalization constant, $\vec{E}_{0 \mu}$, which is obtained with the condition
of having one photon of energy $\hbar \omega$ in the quantization volume $V$, so that 
$|\vec{E}_{0 \mu}|^2/(8 \pi)=\hbar \omega/V$. Next, we perform the sum over photon states as
\begin{equation}
\frac{1}{V} \sum_{photons} \rightarrow 2\int \frac{d^3 \vec{k}_{ph}}{(2\pi)^3}=2\times \frac{4\pi}{(2 \pi c)^3}\int d\omega\, \omega^2,
\label{photonsum}
\end{equation}
where the factor of 2 comes from polarization, $ \vec{k}_{ph}$ is the photon wave vector and the last identity is appropriate for the problem of full spherical symmetry 
(no preferential direction in space) in vacuum that we address in this work. Substituting eqs.~(\ref{M'}-\ref{photonsum}) into eq.~(\ref{golden0}) 
and collecting all the prefactors we obtain
\begin{equation}
\frac{1}{\tau_{ph}}= \frac{4e^2}{m_{e}^{* 2} c^3}\int_{0}^{\infty} d(\hbar \omega)\, \hbar \omega \sum_{i} 
f(e_i) |\tilde{M}_{h,i}|^2 \delta(e_h-e_i+\hbar \omega),
\label{golden}
\end{equation}
 with the matrix elements given simply by
\begin{equation}
\tilde{M}_{h,i}=\langle \Psi_h|\vec{\nabla}\cdot \vec{E}^{*}+2\vec{E}^{*}\cdot \vec{\nabla}|\Psi_i \rangle,
\label{M}
\end{equation}
where in a time-reverse picture, the electric field $\vec{E}$ is the total one inside the sphere corresponding 
to an outgoing electromagnetic plane wave of unit amplitude. 

From eq.~(\ref{golden}) we obtain the visible luminescence spectrum as 
the number of emitted photons with energies between $\hbar \omega$ and $\hbar (\omega+d\omega)$ per unit time and per photon incident onto the sphere as
\begin{equation}
\frac{d I_{VL}}{d(\hbar \omega)}=\frac{\sigma_{abs}(\omega_{in}, R)}{2 \pi R^2}\times  \frac{4e^2}{m_{e}^{* 2} c^3}\hbar \omega \sum_{i}
f(e_i) |\tilde{M}_{h,i}|^2 \delta(e_h-e_i+\hbar \omega),
\label{PL}
\end{equation} 
the first factor on the right hand side of eq.~(\ref{PL}) being the number of holes per photon incident onto the nanosphere. 
Note here that the assumption that all holes are deeply trapped only affects the intensity of the signal but not its spectrum nor the value of
the radiative decay time, both given by eq.~(\ref{golden}). 
We next give details of how we calculate each of the terms in eqs.~(\ref{M}) and (\ref{PL}).

The normalized electric field inside the nanoparticle induced by an x-polarized plane wave, can be written in the dipole approximation, due to the small sizes of interest here, as
\begin{equation}
\vec{E}(r,\theta, \varphi)=
\left(1-\frac{\alpha(\omega,R)}{R^3}\right)(\sin\theta\cos\varphi \hat{i}_r +\cos\theta\cos\varphi \hat{i}_{\theta}-\sin\varphi \hat{i}_{\varphi}),
\label{E} 
\end{equation}
in spherical coordinates, with $\hat{i}_r$, $\hat{i}_{\theta}$ and $\hat{i}_{\varphi}$ being the unitary vectors. 
The polarizability of the sphere $\alpha(\omega, R)$ enters into the evaluation of $\vec{E}$.
We assume it is set by these electrons in the conduction band that can participate in the emission event (i.e. those not scavenged by
oxygen) and, accordingly, it is given by 
\cite{PS_26_113_apell, JPCL_6_1847_carmina, JPCC_120_5074_carmina}
\begin{equation}
\alpha(\omega,R)/R^3= \frac{\epsilon(\omega)-\epsilon_m+2\frac{d_{\theta}(\omega,R)}{R}}
{\epsilon(\omega)+2\epsilon_m+2 \frac{d_{\theta}(\omega,R)}{R}},
\label{alpha}
\end{equation}
where $\epsilon(\omega)$ is the local permittivity of ZnO, 
\begin{equation}
\epsilon(\omega)=\epsilon_{\infty}-\frac{\omega_p^2}{\omega^2-(\frac{\Delta}{\hbar})^2+i\omega\gamma_b},
\label{epsilonL}
\end{equation} 
$\epsilon_m$ is the permittivity of the medium surrounding the sphere (assumed to be frequency independent),
$\epsilon_{\infty}$ is the high-frequency permittivity of ZnO, $\omega_p$ is the plasma frequency with $\omega_p^2=\frac{4 \pi n_e e^2}{m_{e}^{*}}$.
$\Delta$ is the average distance between the conduction band levels defined as \cite{NJP_15_083044_carmina}
\begin{equation}
\Delta= \hbar \omega_p \frac{R_0}{R}, 
\label{gap}
\end{equation}
where  $R_0$ can be estimated from a simple model \cite{SovPhysJETP_21_940_gorkov} as  
$R_{0}=\sqrt{\frac{3\pi a_0}{4m_e^{*} k_F}}$, with $m_{e}^{*}$ being the effective electron mass 
(in units of the electron mass), $a_0$ being the Bohr radius and 
$k_F=(3 \pi^2 n_e)^\frac{1}{3}$ . The 
the complex length $d_{\theta}(\omega,R)$ takes into account the effects of diffuse surface scattering of the electrons\cite{JPCL_6_1847_carmina, JPCC_120_5074_carmina}, which is significant in low charge carrier density systems such as ZnO here considered. 

The electronic states appearing in eq.~(\ref{M}) should be orthonormal because they should be eigenfunctions of the same Hamiltonian.
We take care of this very important fact in an approximate way. 
We start by considering the conduction band electrons as free-electrons confined by an infinite potential well at the surface of the nanosphere.
These states are orthonormal and described by $|\phi_{lnm}\rangle$ for quantum numbers  $(lnm)$, where $(lm)$
are the angular momentum quantum numbers and $n$ quantizes the energy for each $l$ (energies $e_{ln}$).
The corresponding wave functions, expressed in spherical coordinates, are
\begin{equation}
\phi_{lnm}(r,\theta,\varphi)=N_{ln} j_{l}\left(\rho_{ln}\frac{r}{R}\right)Y_{l}^{m}(\theta,\varphi),
\label{phi_lnm}
\end{equation}
were $N_{ln}$ is the normalization constant, $\rho_{ln}$ is the n-order zero of the spherical Bessel function $j_l(\rho)$ and
$Y_{l}^{m}(\theta,\varphi)$ is the spherical harmonic. The corresponding eigenenergies (measured with respect to the bottom of the conduction band) are
\begin{equation} 
e_{ln}=\frac{\hbar^2 \rho_{ln}^2}{2 m_{e}^{*} R^2}.
\label{eln}
\end{equation} 

The nature of the trap is still under debate and several proposals can be found in the literature such as 
Cu impurities\cite{PRL_23_579_dingle},  Zn vacancies \cite{PRB_76_165202_janotti} or oxygen vacancies \cite{JPCB_104_1715_vanDijken} among others. In the case of colloidal QDs, 
the chemical nature of surface ligands is known to play a role in their luminescent properties \cite{JPCB_109_20810_norberg}.
Therefore, in order to keep the problem as simple as possible, and since the energy of the deep hole is approximately in the middle of the band gap, we describe the deep hole state by the hydrogenic wave function
\begin{equation}
\phi_h(\vec{r})=\frac{\alpha_h^{3/2}}{\sqrt \pi}e^{-\alpha_h |\vec{r}-\vec{a}|},
\label{phi_h}
\end{equation} 
 with $\alpha_h=\sqrt{2 m_h e_b/ \hbar ^{2}}$, $m_h$ being the hole mass and  $e_b$ being its binding energy. When
$e_h$ is referred to the bottom of the conduction band, like $e_{ln}$, $e_b=E_g+e_h$, $E_g$ being the band gap of ZnO.
The deep hole is localized at a point $\vec{a}$ inside the sphere that we will take as a parameter. Obviously, this hole state is not, in general, orthogonal to the conduction band states of eq.~(\ref{phi_lnm}). Next, we proceed to construct an orthonormal set of orbitals  $\{\Psi_{\alpha}\}$ from the non-orthogonal set 
$\{\phi_{\beta}\}_{\{\beta=h,(lnm)\}}$ 
by means of the symmetrical orthogonalization procedure proposed by L\"owdin \cite{JPC_18_365_lowdin}:  
\begin{equation}
\Psi_{\alpha}=\sum_{\beta} (\bf{S}^{-1/2})_{\alpha, \beta}\phi_{\beta},
\label{lowdin}
\end{equation} 
where $\bf{S}$ is the overlap matrix with matrix elements $S_{\alpha, \beta}=\langle\phi_{\alpha}|\phi_{\beta}\rangle$.  
As a further approximation we assume all the off-diagonal matrix elements of $\bf{S}$ to be small 
and expand $\bf{S^{-1/2}}$ in powers of the overlap integrals.  Up to order $\mathcal{O}S_{\alpha, \beta}^2$ we get
\begin{equation}
(\bf{S}^{-1/2})_{\alpha, \beta}\approx \delta_{\alpha, \beta}-\frac{1}{2}S_{\alpha, \beta},
\label{S-approx}
\end{equation}
this approximation being especially useful for $a \neq 0$ because it makes possible to obtain an analytical expression for eqs.~(\ref{golden}) and
(\ref{M}).
Using it, the matrix elements $\tilde{M}_{h,i}$  of eq.~(\ref{M}) between orthogonal orbitals can be easily expressed in terms of the corresponding ones
for non-orthogonal orbitals as
\begin{equation}
\tilde{M}_{h,(lnm)}=M_{h,(lnm)}^{n-o}-\frac{1}{2}\sum_{l'n'm'} S_{h,(l'n'm')}M_{(l'n'm'),(lnm)}^{n-o},
\label{Mor}
\end{equation}
with the superscript $n$--$o$ indicating that eq.~(\ref{M}) is to be evaluated between the non-orthogonal states of eqs.~(\ref{phi_lnm}) and (\ref{phi_h}). The details of this calculation are given in the Supplemental Information.

The derivation above strictly  holds for ideally spherical nanoparticles. 
For real spheres, however, because of the unavoidable existence of imperfections in their shape and morphology, 
it is nearly impossible to calculate a detailed level distribution.
A common approach has been to assume the level distribution to be completely random.
\cite{SovPhysJETP_21_940_gorkov,JPCL_8_524_carmina}. On the other hand, the hole state has a large width. It is known that electron-hole recombination at
a defect site results in large reorganization of the local charge and, consequently, strong vibronic transitions. We take
care of these effects by making use of the identity
\begin{equation}
\delta(e_{h}-e_{ln}+\hbar \omega)=\int d e\, \delta(e-e_{ln}) \int d e'\, \delta(e'-e_{h}) \delta (e'-e +\hbar \omega),
\label{delta}
\end{equation}
and then substituting the function $\delta(e-e_{ln})$  by a gaussian distribution of probability function as \cite{JPCL_8_524_carmina}
\begin{equation}
\delta(e-e_{ln})\rightarrow G(e-e_{ln})=\frac{1}{\sigma \sqrt{2\pi}}\exp\left[-\frac{(e-e_{ln})^2}{2 \sigma^2}\right],
\label{gausse}
\end{equation}
for the conduction band states and likewise for the deep hole state
\begin{equation}
\delta(e'-e_{h})\rightarrow G(e'-e_{h})=\frac{1}{\sigma_h \sqrt{2\pi}}\exp\left[-\frac{(e'-e_{h})^2}{2 \sigma_h^2}\right].
\label{gaussh}
\end{equation}
 An appropriate value for $\sigma$ is $\sigma=\Delta/2$, \cite{JPCL_8_524_carmina} $\Delta$ being the average distance 
between the conduction band levels  near 
their Fermi energy defined by eq.~(\ref{gap}). We take $\sigma_h \approx 0.1-0.2$ eV, a typical value for vibrational energies.
However, we assume for simplicity that the matrix elements can be calculated by means of eq.~(\ref{Mor}), which implies that matrix elements 
are not very sensitive to changes in nanoparticle radii by $\approx 7\%$ \cite{JPCL_8_524_carmina} or to changes in the wave function  of the deeply trapped hole.

\section*{Results and discussion}
 
 We perform calculations for the case of  ZnO nanospheres in ethanol using the values:  
$E_g=3.44$ eV, $m_e^{*}=0.28 m_e$, $\gamma_b=$0.1 eV, $\epsilon_{\infty}=3.72$ and $\epsilon_m=1.85$.
The parameters for the deep hole are $m_h=m_e$ and $\epsilon_h=-1.95$ eV (measured with respect to the bottom of the conduction band) 
taken from experiments \cite{JLumin_90_123_vanDijken}. 
In all the calculations to be presented in this work, we include in eq.~(\ref{Mor}) states $(l'n'm')$ with $n' \leq 7$ for each of the  quantum numbers 
$(l'm')$ and this is enough for obtaining converged results. 
In the following, states $(ln)$ will be denoted according to the standard notation as $nS$ for $l=0$, $nP$ for $l=1$, $nD$ for $l=2$ etc. 

\begin{figure} 
\centering
\includegraphics{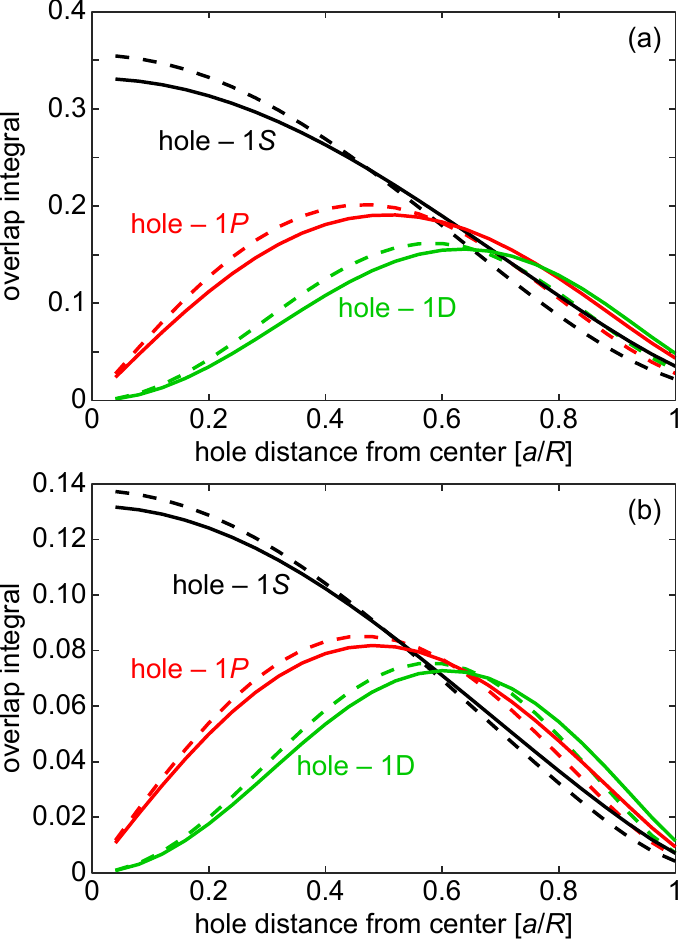}
\caption{The overlap integrals between the deep hole at position $a/R$ from the QD center and the $1S$ (black lines), $1P$ (red lines) and $1D$ (green lines) states of the conduction band electrons as a function of
the distance of the hole to the surface of the sphere. Continuous lines: 
step potential barrier of height $V_0=$ 4 eV, dashed lines: infinite barrier at the surface of the nanoparticles. (a) $R=2$ nm; (b) $R=4$ nm.}
\label{fig1}
\end{figure}

To check the accuracy of eq.~(\ref{S-approx}) we present in
Figure~1a,b the overlap integrals $S_{h,(lnm)}$ for  $l=0,1,2$, $n=1$ and $m=0$, as a function of the distance $a$ of the deep hole to the sphere center, for
$R=2$ nm and $R=4$ nm respectively. We note that the overlap of the deep hole with the $nS$  states of the conduction band electrons is $\approx 0.4$ near the center and 
then decreases as $ a $ approaches the sphere surface. Therefore, while the approximation of eq.~(\ref{S-approx}) is a good one near the surface, it might  not be so good near
the sphere center, especially for the smallest size. For $a=0$ it is not difficult to obtain corrections to eq.~(\ref{S-approx}) up to the order 
$\mathcal{O}S_{\alpha, \beta}^4$ 
and we have checked that this does not change the order of magnitude of the results. Nevertheless, we restrict our calculations to $a \ge 0.2 R$. 
Figure~1 also shows a comparison of the overlap integrals assuming that the conduction electrons are confined 
by an infinite barrier (dashed lines) or by a step potential barrier of height $V_0=4$ eV (continuous lines)
\cite{JPCL_8_524_carmina,PRB_87_155311_schlesinger}. 
The small changes we see motivate our use of infinite barrier model, that is much simpler for analytical and
computational purposes.

\begin{figure} 
\centering
\includegraphics{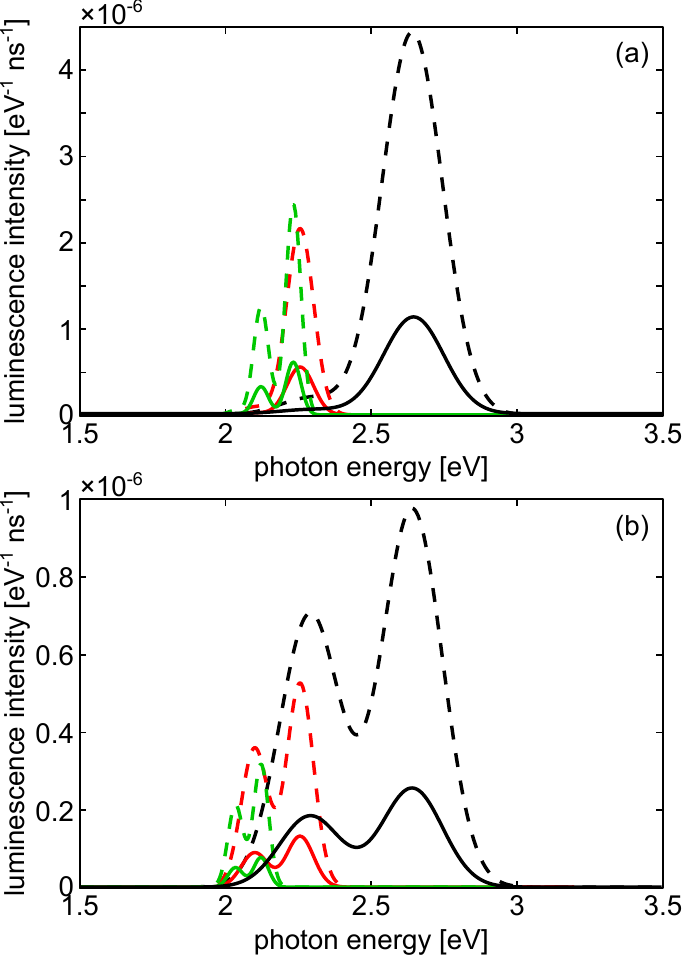}
\caption{Visible luminescence spectra of nanospheres of $R=2$ nm (black lines), $R=3$ nm (red lines) and $R=4$ nm (green lines), 
all having $N_e=4$ electrons in the conduction band, 
calculated using orthogonal (continuous lines) and non-orthogonal (dashed lines) orbitals and $\sigma_h=0$. The hole is localized in (a) at $a=0.2 R$ and in (b) at $a=0.8 R$. A proper orthogonalization of the involved states decreases the emission probabilities and, consequently, increases the radiative
lifetimes, by a factor of 4. Only $1P$ states can contribute to radiative decay of the hole in (a) while $1S$ states can also do it in (b). Therefore the 
shape of the emission band depends on the hole position.}
\label{fig2}
\end{figure}

We first present results under the assumption (a) that the number of electrons in the conduction band is a small number independent of particle size.
To demonstrate the importance
of orthogonalization effects, Figure~2a,b shows calculations of the visible luminescence spectra, using orthogonal (continuous lines)
and non-orthogonal (dashed lines) states, for $N_e=4$ (2 electrons in the $1S$ state and 2 electrons in the $1P$ state),  $R=2, 3$ and 4 nm, $\sigma_h=0$  and for 
$a=0.2 R$ in Figure~2a and $a=0.8 R$ in Figure~2b. 
Note that the non-orthogonal calculation overestimates the spectra by roughly a factor of 4 and, consequently, the radiative lifetimes 
would be underestimated by the same factor.
The peaks in Figure~2a originate from the radiative recombination of the electron in the $1P$ state with the trapped hole of the same spin. The peak intensity decreases 
and its position shifts to lower energy with increasing size, as in the experiments. The shift to lower energy with increasing size is predicted by eq.~(\ref{eln}) for the eigenenergies. 
The decrease in intensity is the consequence of the decrease in the overlap between electron and hole states with increasing size, compare Figure~1a and b, and the matrix elements
appearing in  eq.~(\ref{Mor}) behave in the same way.  Hence this simple model reproduces the experimental 
behavior of the evolution of position and intensity of the steady-state green luminescence with particle size at room temperature 
\cite{ChemPhysLett_122_507_koch, JPhysChem_91_3789_bahnemann, JPCB_104_1715_vanDijken, JLumin_90_123_vanDijken}.
The small shoulder to the left of the peak 
in Figure~2a originates from the radiative recombination of the electron in the $1S$ state with the trapped hole of the same spin. 
This process, that is strictly forbidden if the hole is localized at $a=0$, shows as a faint feature for small values of $a$ but gets important  
if the deep hole is localized near the nanoparticle surface, 
as can be appreciated in Figure~2b, where both peaks are of nearly the same intensity. Nevertheless the relative intensity of both peaks does not change much for $a \ge 0.5 R$

\begin{figure} 
\centering
\includegraphics{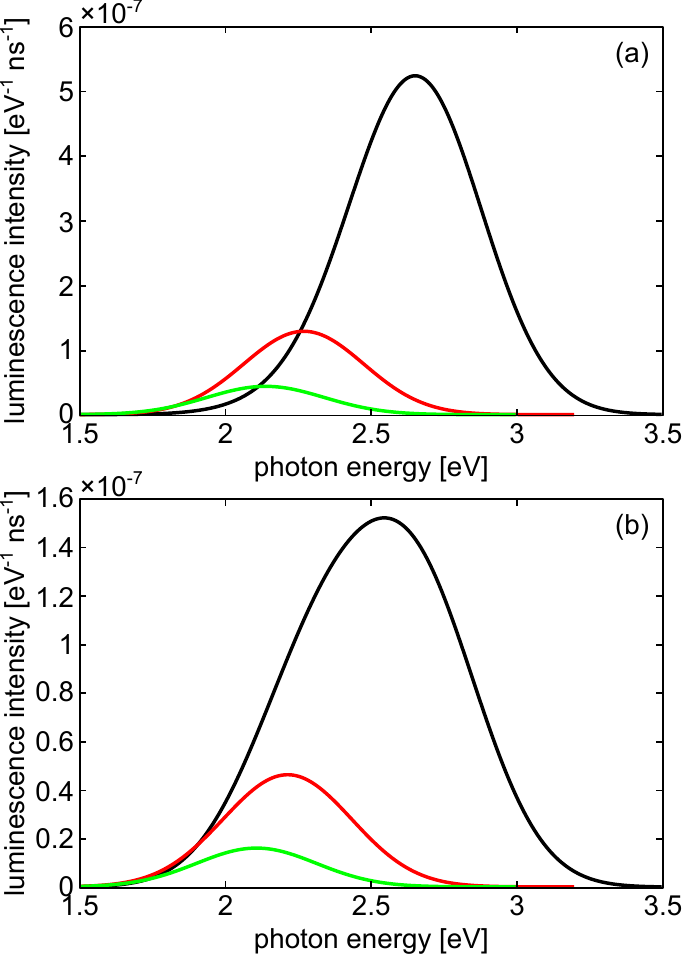}
\caption{Visible luminescence spectra of nanospheres of $R=2$ nm (black lines), $R=3$ nm (red lines) and $R=4$ nm (green lines), 
all having $N_e=4$ electrons in the conduction band, assuming
the hole to be localized in (a) at $a=0.2 R$ and in (b) and $a=0.8 R$ with an energy broadening of the hole 
$\sigma_h=0.2$ eV. The emission bands of Figure 2 evolve into broad continuous bands, with both the maximum intensity and its energetic position decreasing with increasing size, in qualitative agreement with experiments.}
\label{fig3}
\end{figure}

The calculations of Figure~2 are for a hole state having zero width.  Figure~3 displays the photoluminescence spectra of eq.~(\ref{PL}) for the same parameters as in Figure~2 but 
including an energy broadening of the deep hole of $\sigma_h=0.2$ eV, where it can be observed how the atomic-like peaks merge and form a continuous broad band. 
Although the behavior with particle size of this band follows the experimental trends mentioned above, 
a more quantitative analysis shows that the relative changes are too quick in comparison with experiments. 
The same happens if we assume to have $N_e=2$ in the QDs (not shown here). Figure~3 of ref.\cite{JPCB_104_1715_vanDijken} shows a quick decrease of the peak intensity with increasing size for radii smaller than ca. 1.5 nm
which can be modeled by the present approximation, but this quick decrease is  followed by almost constant values of the intensity for radii increasing up to 3.5 nm, 
while a continuous decrease is obtained in the calculations.  The quick decrease in peak energy with increasing
radius is also not seen in the experiments\cite{JPCB_104_1715_vanDijken}, 
where this energy decreases from ca. 2.4 eV for the smallest size to ca. 2.2 eV for the larger radius. 

\begin{figure}  
\centering
\includegraphics{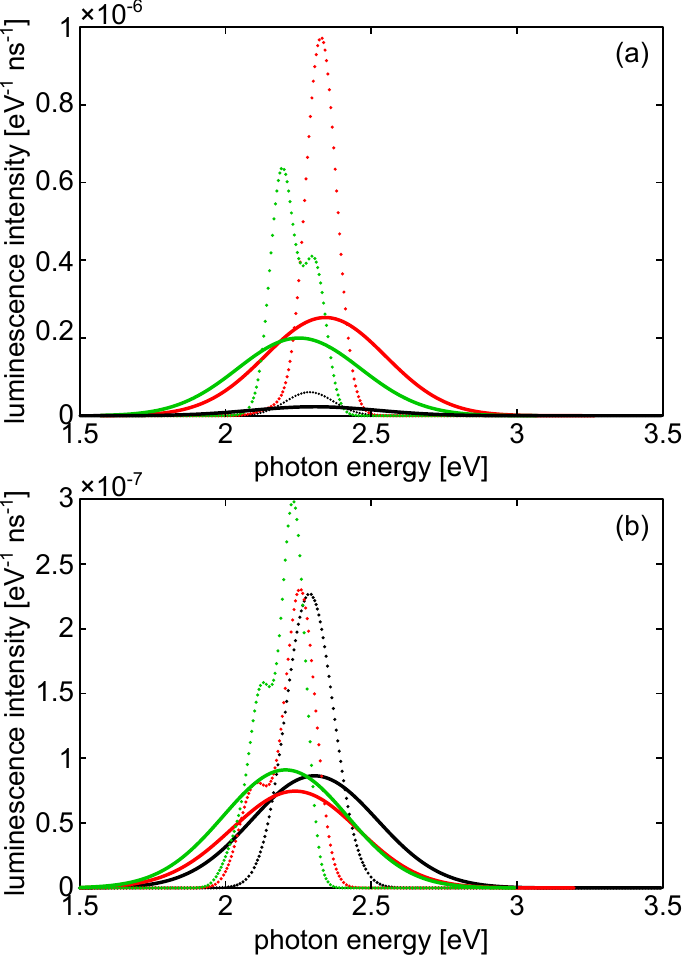}
\caption{Visible luminescence spectra of nanoparticles of $R=2$ nm (black lines), $R=3$ nm (red lines) and $R=4$ nm (green lines), 
all having the same density $n_e=0.6 \times 10^{20}$ cm$^{-3}$ of electrons in the conduction band, assuming
the hole to be localized in (a) at $a=0.2 R$ and in (b) at $a=0.8 R$, 
for $\sigma_h=0$ (dotted lines) and $\sigma_h=0.2$ eV (continuous lines). The evolution of the intensity of the
emission bands with QD size is very different depending on where inside the nanoparticle is the hole localized. A better agreement with the experiments at room temperature \cite{JPCB_104_1715_vanDijken}
is obtained assuming the hole is located near the surface, as in (b).}
\label{fig4}
\end{figure}

We then turn to our assumption (b) and consider that the density of conduction band electrons $n_e$ is independent of particle size. This is motivated
by the fact that well characterized ZnO colloidal QDs in toluene, with sizes ranging from 2 to 6 nm in radii, can be charged, under anaerobic conditions, all to the same maximum electron density of $n_e= (1.4 \pm 0.4) \times 10^{20}$ cm$^{-3}$ \cite{ACSNano_8_1065_schimpf}
and in previous works we used the lower limit of this density 
to reproduce the IR absorption properties of these QDs \cite{JPCL_6_1847_carmina, JPCC_120_5074_carmina,JPCL_8_524_carmina}. 
Guided by these maximum values, we consider the aerated QDs to have approximately one half of that density. Actually, 
since in our calculations we always have an even number of conduction electrons (half of each spin) and the minimum value of $N_e$ is 2, the minimum density we consider is 
$n_e=0.6  \times 10^{20}$ cm$^{-3}$ corresponding to 2 electrons in a sphere of $R=2$ nm.  Under the assumption that all QDs have 
this same density of electrons in the conduction band, a sphere of $R=3$ nm has $N_e=6$ electrons and a sphere of $R=4$ nm has $N_e=16$ electrons. 

The visible photoluminescence spectra for equal electron density are presented in Figure~4a,b, for $a=0.2 R$ and $a=0.8 R$ respectively.  
Looking at the results for $\sigma_h=0$ (dots), we only see one small peak at $\omega \simeq 2.3$ eV for $R=2$ nm which originates from the electron in the $1S$ level. 
As we said above, this transition would be strictly forbidden if the hole was localized at the sphere center. Consequently, this feature is small for $a=0.2 R$ but not for
$a=0.8 R$. For $R=3$ nm, the prominent peak at 
$\omega \simeq 2.25$~eV in Figure~4 originates from the radiative decay of electrons in the $1P$ level while the shoulder at $\omega \simeq 2.1$ eV 
originates from electrons in the $1S$ level. For $R=4$ nm the conduction band levels $1S$, $1P$ and $1D$ are populated with electrons and the corresponding features are
seen in the photoluminesce spectra. Therefore, as size increases, the energy levels move down in energy but simultaneously higher energy levels become populated
with electrons with the net effect being a slow variation of the position of the maximum with size.
In fact, when we give a line broadening to the hole, the atomic-like features merge into broad bands with maxima at energy positions
that decrease with increasing radius as before but at a pace more in accordance with experiments. The values of these maxima in Figure~4b are 
$\omega_{max} \simeq 2.31$ eV, 
 $\omega_{max} \simeq 2.24$ eV, and $\omega_{max} \simeq 2.21$ eV, for $R=2$, 3 and 4 nm, respectively. With respect to the dependence of the maximum intensity on size, we note 
 in Figure~4a that the peak of the  $R=2$ nm nanosphere is too small for $a=0.2 R$, and the experimental characteristics of a weak dependence of intensity on size 
 in the range of sizes 1.5~nm $\le R \le$ 3.5~nm is best
 reproduced if we assume the deeply trapped hole to be localized near the surface. This weak dependence is the net effect of the competition between 
decreasing overlap and increasing population of high energy levels with increasing size. 
Therefore, these calculations confirm the conclusions of van Dijken et al.
\cite{JPCB_104_1715_vanDijken} showing
evidence of fast surface trapping of holes in the trap emission process. It is also experimentally known that passivation of surface defects quenches the visible photoluminescence \cite{JPCL_4_3024_faucheux, JPCB_109_20810_norberg}.
 The values of the radiative lifetimes for the cases in Figure~4b are 
 $\tau_{ph}= 0.53$~$\mu$s, 0.89~$\mu$s and 0.98~$\mu$s for $R= 2$, 3 and 4 nm respectively, increasing with increasing size as in the experiment,
 see Figure~4 of ref.\cite{JPCB_104_1715_vanDijken}.
Moreover, a value of $\tau= 1.34$~$\mu$s was measured for $R=3$ nm and $\tau= 0.92$~$\mu$s for $R=1$ nm at room temperature, both 
comparable to our calculation.
 
\begin{figure} 
\centering
\includegraphics{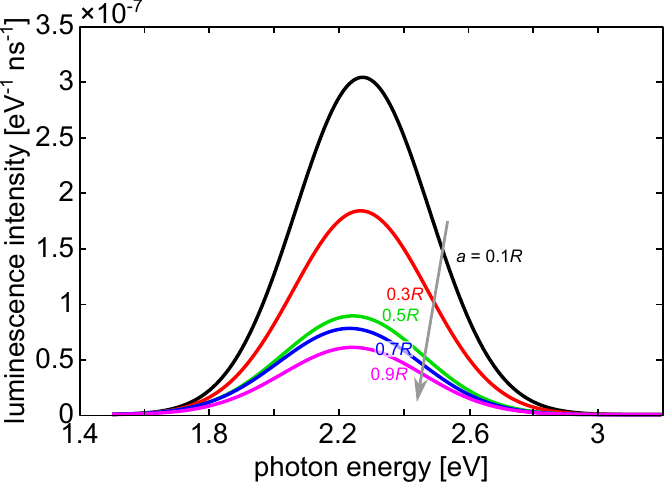}
\caption{Visible luminescence spectra of nanoparticles of  $R=3$ nm with an electron density of $n_e=0.6 \times 10^{20}$ cm$^{-3}$, $\sigma_h=0.2$ eV, for an increasing hole -- QD-center distance $a$. 
 The intensity decreases and the radiative lifetime increases (both with increasing $a$) by approximately the same factor 
as the corresponding experimental magnitudes do when the temperature is increased from 4 K to room temperature. This suggests that 
the holes are predominantly trapped at the particle center at low temperatures but at the particle surface at room temperature.}
\label{fig5}
\end{figure}

 Based on the results presented above, we now give a possible reason for the experimentally observed
 behavior of  the visible luminescence spectra and radiative lifetimes of the aerated QD with temperature presented in Figures~5B and 6 of van Dijken et al.\cite{JPCB_104_1715_vanDijken} for nanoparticles of $R=3$ nm. 
 Figure~5 displays calculated spectra for nanospheres of $R=3$ nm for different values of the position of the hole within the particle.
 It is interesting to note the fast decrease of the maximum intensity as
the hole moves from the center to the surface, the factor of 5 difference we find here being reminiscent of the same fast decrease of 
the experimental intensity with increasing temperature shown in Figure~5B of van Dijken et al.\cite{JPCB_104_1715_vanDijken} 
Furthermore, the decay of the visible emission signal is found to be single exponential at very low and at room temperatures with values 
of the decay time of 0.275~$\mu$s and 1.34~$\mu$s at $T=4$ K and room temperature, respectively. 
The theoretical values of the radiative lifetime are 0.24~$\mu$s and 1.14~$\mu$s for $a=0.1 R$ and $a=0.9 R$, respectively, and are comparable to the experimental values. These results suggest that the holes are predominantly trapped in the bulk of the nanoparticle at low temperatures and at the surface at
room temperature. So increasing the temperature increases the probability for the holes to be transferred to defects at the surface of the nanoparticles.
It is interesting to note here that an exponential decay of the visible emission signal is also seen in \emph{single crystals} at low temperatures 
(smaller than 20 K) with $\tau \simeq 0.44$~$\mu$s \cite{PRL_23_579_dingle}, a value similar to the one seen for $R=3$ nm nanoparticles at $T=4$~K quoted above. 
The transition from low to room temperatures seen by van Dijken et al.\cite{JPCB_104_1715_vanDijken} is at $T \sim 100-125$ K or 
$k_{B}T \simeq 0.01$ eV. 
 The binding energy of the hole can depend on whether it is trapped at the center or at the surface of the nanoparticle and a difference
in binding energies on the order of 0.01~eV is enough to thermally populate/depopulate one or the other site at that transition temperature. 
In general, there would be a spatial distribution of traps within the 
nanoparticle leading to a multiexponential decay of the luminescent signal with exponents than depend on the site where the trap is localized. 
The calculations of Figure~5, however, differ from the experimental results in that the theoretical 
position of the maximum moves to lower energies by 0.05~eV 
at most while differences of ca. 0.1~eV are found in the experiment, see Figure~5B of van Dijken et al.\cite{JPCB_104_1715_vanDijken}. 
It could be that the interaction of the localized hole with the vibronic modes, which is certainly temperature dependent, 
also affects its energetic position, an important effect not considered in the present work.

\begin{figure} 
\centering
\includegraphics{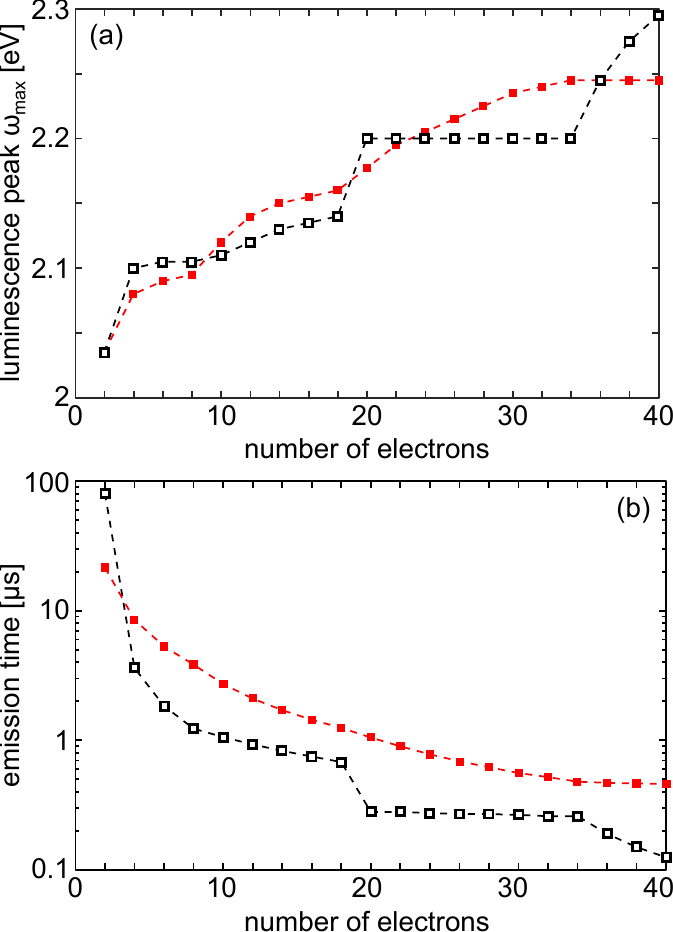}
\caption{(a) Position in energy of the maximum of the visible luminescence spectra and (b) radiative time as a function
of the number of conduction band electrons for ZnO nanospheres of $R=4.5$ nm in toluene for $a=0.2 R$ (black open squares) and $a=0.8 R$ (red full squares).
While the maximum energy blue-shifts with increasing number of electrons by $\approx$ 10$\%$ the radiative time strongly decreases and only tends to saturate at large
values of $N_e$.}
\label{fig6}
\end{figure}

Very recently it has been verified experimentally that air-free QDs of different sizes that are photodopped to the same dopping level,
all reach the same average value of their electronic density independent of size, at $n_e \gtrsim 0.2 \times 10^{20}$ cm$^{-3}$ \cite{JACS_137_11163_carroll}. 
The better agreement between theory and experiment, which we find under the assumption that the density of conduction band electrons is independent of size,
implies that this is also the case in the presence of air.
This is in accordance with other experimental observations such as, for example,  Cohn et al. 
\cite{JPCC_116_20633_cohn} who \emph{``conclude a surprisingly high probability of exciting unintentionally charged ZnO nanocrystals even
under aerobic conditions and even without deliberate addition of hole scavengers''} and other works \cite{JPCB_104_4355_vanDijken,JPCC_114_220_stroyuk}
which find that it is more easy to charge with electrons the biggest crystals than the smaller ones. Hence it is important to analyze the dependence of 
the visible emission characteristics with electron density and we do it by considering a system of ZnO nanoparticles in toluene ($\epsilon_m=2.25$), where it is known that they can be charged up to a maximum electron density of $n_e \approx 1 \times 10^{20}$ cm$^{-3}$ \cite{ACSNano_8_1065_schimpf}.
As in the work of Cohn et al. \cite{JPCC_116_20633_cohn} we consider spheres of $R=4.5$ nm and present our results for $\omega_{max}$ and $\tau_{ph}$ as a function of $N_e$
in Figure~6a,b, respectively. In this case, the maximum density is obtained for $N_e=36$ electrons.
As the number of electrons increases, the maximum of the spectra in Figure~6a blue shifts by 0.20--0.25 eV, that is $\approx$10$\%$ at most, in accordance with other findings  \cite{JPCB_104_4355_vanDijken, JPCC_115_21635_yamamoto, JPCL_4_3024_faucheux}. 
The jumps in the value and/or slope of $\omega_{max}$ seen at $N_e=4, 10, 20$ and 36 occur when a new shell starts to be filled.
In contrast to $\omega_{max}$, the radiative time shows a pronounced decrease for small $N_e$ and tends to saturate for larger values.
Experiments by Cohn et al. \cite{JPCC_116_20633_cohn} for uncharged crystals report a multiexponential decay that is fitted reasonably well 
by two time constants, a fast one of $\tau_1 \approx 0.2$~$\mu$s and a slow one of $\tau_2 \approx 1.8$~$\mu$s, with the slow component
constituting $\approx 90 \%$ of the amplitude. Assigning this component to the radiative decay of holes localized near the surface, that
value of $\tau_2$ 
is obtained in our calculation for $N_e \approx 14 $ electrons, or $n_e \approx 0.4 \times 10^{20}$ cm$^{-3}$,
a value of the electron density similar to the one found above for reproducing experiments of aerated ZnO QDs in ethanol. 
However, the calculated radiative lifetime decreases continuously when increasing $N_e$ above this value which   
 seems to be in contradiction with the observation of Cohn et al.\cite{JPCC_116_20633_cohn} who report that the slow component is weakly affected by the addition of extra electrons, but the effect is not quantified. 
Nevertheless, it is known that the accumulation of electrons in the conduction band strongly perturbs the dynamics of the trapped holes 
\cite{ChemPhysLett_122_507_koch, JPhysChem_91_3789_bahnemann, JPCB_104_1715_vanDijken, JLumin_90_123_vanDijken, JPCB_104_4355_vanDijken, JPCC_114_220_stroyuk, JPCC_115_21635_yamamoto,
JPCC_116_20633_cohn,JPCL_4_3024_faucheux} and, consequently, their decay time will be equally perturbed. 
Therefore, although our simple theoretical model is able to reproduce many experimental trends, 
we conclude that improvement of the theory as well as a better experimental characterization of the charge and distribution of defects in the nanocrystals is essential to fully understand and control the luminescent properties of ZnO QDs.

\section*{Conclusions}

In this work we have developed a theory for the photoluminescence of ZnO QDs using Fermi's golden rule together with a simple model for the electronic states. 
We assumed the green luminescence to be generated in the
radiative recombination of a delocalized conduction band electron with a hole trapped within the particle and analyzed the dependence of 
the spectral characteristics on the particle size and spatial localization of the hole.
The behavior of this band with particle size can be understood in terms of quantum size effects of the electronic states and their overlap 
with the localized hole.
Focusing the comparison of our results with the detailed experiments performed for ZnO nanoparticles
in ethanol under aerobic conditions\cite{JPCB_104_1715_vanDijken, JLumin_90_123_vanDijken}, we conclude that the experimental trends 
found at room temperature can be accurately reproduced by assuming that the hole is located at the surface of the QD. 
Based on our results of the dependence of emission intensity on the location of the hole,
we suggest that the anomalous dependence of the decay time with temperature found by van Dijken et al. \cite{JPCB_104_1715_vanDijken} for $R=3$ nm
can be the effect of an increasing probability for the holes to be trapped at the surface with increasing temperature.
Furthermore, the calculated values of the radiative lifetimes are comparable to the experimental values of the decay time of the 
visible emission signal.
We also studied the visible emission band as a function of the number of conduction band electrons in the nanoparticle 
finding a pronounced dependence of the radiative lifetime
that does not seem to conform to experiment. We thus conclude that further improvement of the theoretical model, as well further experimental effort
providing a better characterization of the distribution of traps and the charge of the nanoparticles, is essential to fully understand and control their emission properties.


\begin{acknowledgement}
RCM acknowledges financial support from the Spanish Ministry of Economy and Competitiveness through  
the Mar\'ia de Maeztu Programme for Units of Excellence in R\&D (MDM-2014-0377) and the project MAT2014-53432-C5-5-R.
TJA thanks the Polish Ministry of Science and Higher Education for support via the Iuventus Plus project IP2014 000473. 
TJA and SPA acknowledge financial support from the Swedish Foundation for Strategic Research via the 
Functional Electromagnetic Metamaterials for Optical Sensing project SSF~RMA~11.
\end{acknowledgement}

\begin{suppinfo}
In the Supporting Information we present full derivations of the calculations of overlap integrals and matrix elements.
\end{suppinfo}


\providecommand*\mcitethebibliography{\thebibliography}
\csname @ifundefined\endcsname{endmcitethebibliography}
  {\let\endmcitethebibliography\endthebibliography}{}

\end{document}